\documentclass
[preprint,showpacs,byrevtex,10pt,twocolumn,tightenlines,prl,letterpaper]{revtex4}%
\usepackage{amsfonts}
\usepackage{amsmath}
\usepackage{amssymb}
\usepackage[dvips]{graphicx}
\usepackage{hyphenat}%
\providecommand{\U}[1]{\protect\rule{.1in}{.1in}}

\begin{document}
\preprint{ }
\title{Conductance relaxation in GeBiTe - slow thermalization in an open quantum system.}
\author{Z. Ovadyahu}
\affiliation{Racah Institute of Physics, The Hebrew University, Jerusalem 91904, Israel }


\begin{abstract}
This work describes the microstructure and transport properties of
GeBi$_{\text{x}}$Te$_{\text{y}}$ films with emphasis on their
out-of-equilibrium behavior. Persistent-photoconductivity (PPC), previously
studied in the phase-change compound GeSb$_{\text{x}}$Te$_{\text{y}}$ is also
quite prominent in this system. Much weaker PPC response is observed in the
pure GeTe compound and when alloying GeTe with either In or Mn. Films made
from these compounds share the same crystallographic structure, the same
p-type conductivity, a similar compositional disorder extending over
mesoscopic scales, and similar mosaic morphology. The enhanced
photo-conductive response exhibited by the Sb and Bi alloys may therefore be
related to their common chemistry. Persistent-photoconductivity is observable
in GeBi$_{\text{x}}$Te$_{\text{y}}$ films at the entire range of sheet
resistances studied in this work ($\approx$10$^{\text{3}}\Omega$ to $\approx
$55M$\Omega$). The excess conductance produced by a brief exposure to infrared
illumination decays with time as a stretched-exponential (Kohlrausch law).
Intrinsic electron-glass effects on the other hand, are observable in thin
films of GeBi$_{\text{x}}$Te$_{\text{y}}$ only for samples that are
strongly-localized just like it was noted with the seven electron-glasses
previously studied. These include a memory-dip which is the defining attribute
of the phenomenon. The memory-dip in GeBi$_{\text{x}}$Te$_{\text{y}}$ is the
widest amongst the germanium-telluride alloys studied to date consistent with
the high carrier-concentration \textit{N}$\geq$10$^{\text{21}}$cm$^{\text{-3}%
}$ of this compound. The thermalization process exhibited in either, the
PPC-state or in the electron-glass regime is sluggish but the temporal law of
the relaxation from the out-of-equilibrium state is distinctly different.
Coexistence of the two phenomena give rise to some non-trivial effects, in
particular, the visibility of the memory-dip is enhanced in the PPC-state. The
relation between this effect and the dependence of the memory-effect magnitude
on the ratio between the interparticle-interaction and quench-disorder is discussed.

\end{abstract}
\maketitle

\section{Introduction}

The approach to equilibrium of quantum systems is a fundamental problem that
has received wide theoretical attention, mostly in close systems \cite{1,2}.
Thermalization of open systems is a less researched subject although it is
relevant for most naturally occurring processes. This is the case for
out-of-equilibrium electronic transport in solids where coupling to a
heat-bath via phonons usually plays a role at any finite temperature. In
metals and semiconductors the rate of electron-phonon (e-ph) inelastic
scattering $\gamma_{\text{in}}$ may be appreciable; even at sub-Kelvin
temperatures $\gamma_{\text{in}}$ is typically 10$^{\text{4}}$-10$^{\text{6}}%
$s$^{\text{-1}}$ at 1K \cite{3}. An efficient e-ph coupling is the main reason
for the fast relaxation of the electronic system after it has been taken out
of equilibrium by a light-pulse or by Joule-heating it.

There are however situations where electronic relaxation is a sluggish process
despite the presence of phonons; persistent photoconductivity (PPC), and the
relaxation exhibited by electron-glasses are examples for such cases. In both
phenomena the electronic conductance G is enhanced in their out of the
equilibrium state, and in both the relaxation of G from the excited state may
be a very slow process that may be monitored over time scales that are many
orders of magnitude longer than phonon relaxation times. The slow relaxation
associated with these phenomena make systems that exhibit them prime
candidates for experimentally studying thermalization in open quantum systems.

PPC has been observed in lightly-doped semiconductors with
carrier-concentration \textit{N} typically smaller than 10$^{\text{18}}%
$cm$^{\text{-3}}$ \cite{4,5,6,7,8}. Recently this phenomenon was observed in
GeSb$_{\text{x}}$Te$_{\text{y}}$ alloys, p-type degenerate semiconductors with
\textit{N} $\approx$10$^{\text{20}}$cm$^{\text{-3 }}$\cite{9}. When Anderson
localized, samples of GeSb$_{\text{x}}$Te$_{\text{y}}$ exhibited both
intrinsic \cite{10} electron-glass effects and PPC. The interplay of the two
coexisting phenomena showed non-trivial effects \cite{10,11}.

In this work we discuss attempts to obtain similar results for other systems
based on GeTe by incorporating either In, Mn, or Bi as the third element in
the alloy. Despite the high concentration of these chemically different
foreign elements, the resulting ternary compounds shared the same
crystallographic structure and p-type conductivity of the GeTe parent
compound. On the other hand, the transport properties differed markedly among
the produced alloys, in particular in terms of their PPC performance.
Incorporating In or Mn atoms in the GeTe matrix did not seem to affect the PPC
of the pure compound. Bismuth however, proved to give enhanced PPC effects
similar to the behavior of the phase-changed material GeSb$_{\text{x}}%
$Te$_{\text{y}}$ \cite{10}. In addition, GeBi$_{\text{x}}$Te$_{\text{y}}$
samples appear to be an efficient system for studying transport effects
associated with the interplay between disorder and interactions: The material
is easy to fabricate, it is flexible in terms of being able to vary its
disorder over a wide range including driving it insulating without making it
granular. The bulk of the work described below is mostly devoted to the
results obtained with this system.

The persistent photo-conductivity and electron-glass features may be
separately observed in GeBi$_{\text{x}}$Te$_{\text{y}}$ samples and exhibit
their distinct relaxation laws. When these phenomena coexist, on the other
hand, the electron-glass features may be significantly modified, in
particular, the memory-dip seems to have a larger magnitude in the PPC-state.
It is argued that the ratio between the quench-disorder and inter-particle
interaction plays a role in the visibility of the memory-dip. All other things
being equal, this ratio is larger when the carrier-concentration \textit{N} of
the electron-glass is higher, which consistently results in a smaller
magnitude of the memory-dip. However, in the PPC-state the
interparticle-interaction to disorder ratio actually increases relative to the
dark-state thus, in turn, enhancing the memory-dip magnitude.

Details of sample preparation, characterization, and their various
measurements techniques are described in the next section. The results and
their discussion are given in section III.

\section{Experimental}

\subsection{Sample preparation and characterization}

Samples used in this study were prepared by co-depositing GeTe and either Bi,
In, or Mn on room temperature substrates in a high-vacuum system (base
pressure 0.8-1x10$^{\text{-7}}$mbar). The GeTe (Equipment Support Company,
USA) was e-gun deposited with rates of 0.6-1\AA /second while Bi, In, or Mn
were evaporated from a Knudsen source with a rate chosen such that the alloy
composition should be close to 1:1:1. Film thickness varied in the range
40-60\AA ~for the films measured in this work. Lateral dimensions of the
samples used for transport measurements were 0.3-0.5mm long and 0.5mm wide.

Two types of substrates were used; 1mm-thick microscope glass-slides, and
1$\mu$m SiO$_{\text{2}}$ layer thermally grown on $\langle$100$\rangle$
silicon wafers. The Si wafers (boron-doped with bulk resistivity $\rho\simeq$
2x10$^{\text{-3}}\Omega$cm) were employed as the gate electrode in\ the
field-effect measurements. The microscope glass-slides were mostly used for
Hall-Effect measurements performed at room-temperatures. Theses revealed
p-type carrier-concentration \textit{N} in all these alloys. For the
GeBi$_{\text{x}}$Te$_{\text{y}}$ films, that were the main system used in this
work, \textit{N} varied in the range of (6-9)x10$^{\text{21}}$cm$^{\text{-3}}$.

Each deposition batch included samples for transport measurements, samples for
Hall-effect measurements, and samples for structural and chemical analysis.
For the latter study, carbon-coated Cu grids were put close to the sample
during its deposition. These grids received the same post-treatment as the
samples used for transport measurements.

Transmission-electron-microscopy (TEM), using the Philips Tecnai G2) were
employed to characterize the films composition and microstructure.

Polycrystalline samples of Ge(M)$_{\text{x}}$Te$_{\text{y}}$ (where M stands
for either Bi, In, or Mn) were obtained by mounting the as-deposited
(amorphous) films on a hot-plate set to$\ $a~temperature T$_{\text{H}}%
$=470$\pm$5K for $\approx$2 minutes during which the sample was crystallized.

A TEM micrograph and associated diffraction pattern of typical GeBi$_{\text{x}%
}$Te$_{\text{y}}$ and GeMn$_{\text{x}}$Te$_{\text{y}}$ films produced in the
above manner are shown in Fig.1 and Fig.2 respectively. These TEM micrographs
and diffraction patterns illustrate the polycrystalline nature of the films
and a tight, space-filling packing of the crystallites. The main difference
between the Bi and the Mn alloys is obviously their different grain-sizes.
These are just few nanometers for the GeBi$_{\text{x}}$Te$_{\text{y}}$ film as
compared with $\approx$100nm for GeMn$_{\text{x}}$Te$_{\text{y}}$. Similar
grain-sizes were observed in our GeIn$_{\text{x}}$Te$_{\text{y}}$ films. The
large disparity in grain-size is also reflected in the diffraction patterns
(Fig.1 and Fig.2). Diffraction patterns taken from these films were consistent
with the rhombohedral (R-3m) phase of GeTe in all samples made of the three
ternary compounds.

Several types of structural defects may be observed in these micrograph; grain
boundaries and twinning being the most prevalent. These defects, as well as
the compositional-disorder (discussed next), and surface scattering are
presumably responsible for restricting the mobility of the films.

The main difference between the GeBi$_{\text{x}}$Te$_{\text{y}}$ and the
GeMn$_{\text{x}}$Te$_{\text{y}}$ and GeIn$_{\text{x}}$Te$_{\text{y}}$ films is
their mobility. For the thickness range of 40-60\AA , the GeBi$_{\text{x}}%
$Te$_{\text{y}}$ samples had sheet resistance R$_{\square}$ in the range
2k$\Omega$-55M$\Omega$ at 4K. With this range it was possible to cover a large
part of the strongly-localized regime (R$_{\square}$%
$>$%
h/e$^{\text{2}}$) which is a pre-requirement for observing inherent
electron-glass effects \cite{12}. In contrast, we were not yet able to push
R$_{\square}$ much above $\approx$10$^{\text{2}}$k$\Omega$ in samples made
from either of the two other alloys even by deliberate surface-oxidation. This
is presumably the reason for our failure to detect electron-glass effects in
GeMn$_{\text{x}}$Te$_{\text{y}}$ and GeIn$_{\text{x}}$Te$_{\text{y}}$ films.
The higher resistance obtainable in the GeBi$_{\text{x}}$Te$_{\text{y}}$ alloy
may be in part due to their much smaller grain-size as noted above. However,
the grain-size in the previously studied GeSb$_{\text{x}}$Te$_{\text{y}}$
\cite{10} was even larger than in GeMn$_{\text{x}}$Te$_{\text{y}}$ and
GeIn$_{\text{x}}$Te$_{\text{y}}$ and had more pronounced texture (which means
less boundary scattering) than that observed in GeMn$_{\text{x}}$%
Te$_{\text{y}}$ while films with R$_{\square}$ $\geq$50M$\Omega$ were easily
produced even in thicker films \cite{10}. Inter-grain scattering is therefore
not likely to be the main source of disorder in these alloys.

The focus of this work is the difference between the two mechanisms for slow
relaxation and therefore, in the following, data are shown mainly for the
GeBi$_{\text{x}}$Te$_{\text{y}}$ samples where both PPC and electron-glass
effects are observable.%
\begin{figure}[ptb]%
\centering
\includegraphics[
height=3.039in,
width=3.039in
]%
{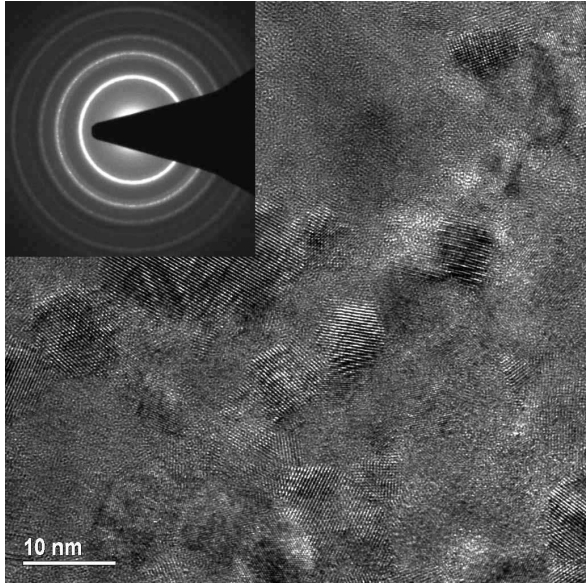}%
\caption{Bright-field of a 5nm film of GeBi$_{\text{x}}$Te$_{\text{y}}$ and an
associated diffraction-pattern. The micrograph shows polycrystalline structure
with grain-size of the order of 1-2nm which ccounts for the rather uniform
rings in the diffraction-pattern.}%
\end{figure}
\begin{figure}[ptb]%
\centering
\includegraphics[
height=3.039in,
width=3.039in
]%
{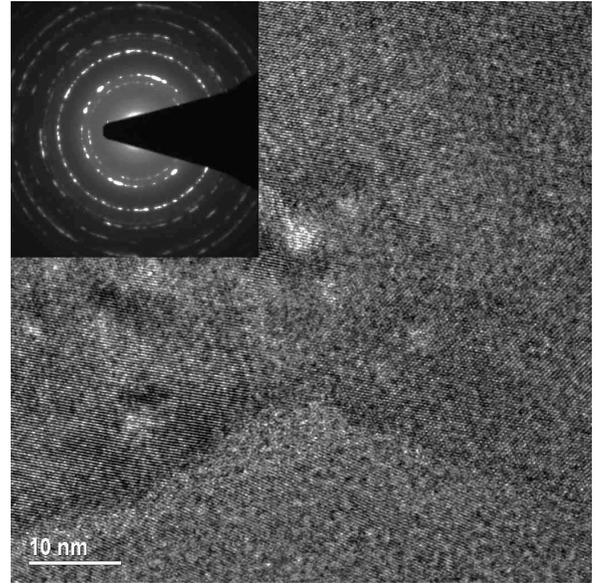}%
\caption{Bright-field of a 5nm film of GeMn$_{\text{x}}$Te$_{\text{y}}$ and an
associated diffraction-pattern taken under similar conditions as the sample in
Fig.1 (same selected-area for the diffraction pattern). The typical grain-size
in this case is clearly much larger than in the GeBi$_{\text{x}}$%
Te$_{\text{y}}$ sample. The relatively large grain-size in this case is also
reflected in the "spotty" rings of the diffraction-pattern.}%
\end{figure}

The stoichiometry of the GeBi$_{\text{x}}$Te$_{\text{y}}$ films was measured
by EDS (Energy Dispersive Spectroscopy) attachment of the TEM. Different
preparation runs produced films that usually had the average stoichiometry
close to the desired goal mentioned above. However, there was noticeable
composition heterogeneity on a mesoscopic spatial scale. This kind of disorder
is quite common in alloys and may be accompanied by spatial fluctuations in
carrier-concentration \cite{13}. A similar composition heterogeneity has been
seen also in GeTe \cite{14}. Figure 3 shows the distribution of local
stoichiometry (on a mesoscopic 40nm scale) across a typical GeBi$_{\text{x}}%
$Te$_{\text{y}}$ film.%
\begin{figure}[ptb]%
\centering
\includegraphics[
height=2.239in,
width=3.039in
]%
{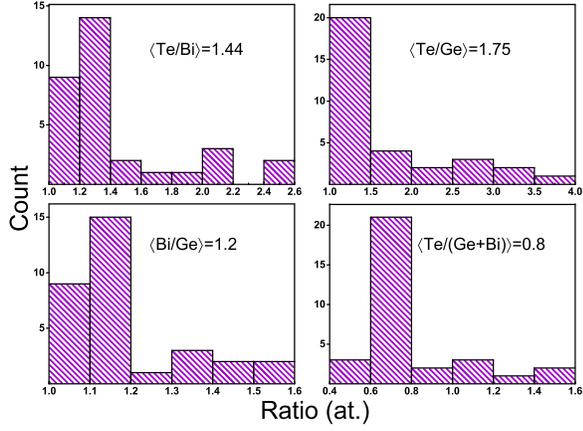}%
\caption{Histograms of the atomic ratios for the three components of the
GeBi$_{\text{x}}$Te$_{\text{y}}$ sample shown in Fig.1 above. The histograms
are based on 35 local EDS measurements across the film, each sampling the
contribution from a 40x40nm$^{\text{2}}$ area. Note that the most likely ratio
is close to a composition of 1:1:1 but there is a significant scatter around
this value. Note that the average ratio for Te/Ge deviates from unity while
Te/(Ge+Bi)$\approx$1 which may hint that the Bi atoms reside on Ge-sites in
the crystal.}%
\end{figure}
Composition variations may be accompanied by local variations of the
carrier-concentration\textit{,} and these are of particular importance in
superconducting materials which are notoriously sensitive to the value of
\textit{N}. In the presence of disorder such inhomogeneities may lead to the
appearance of superconducting islands embedded in an insulating matrix making
it effectively a granular system \cite{13}. One should be aware of these
non-uniform structural aspects whenever the transport property one measures
has a spatial scale that is smaller or comparable with the scale of the inhomogeneities.

\subsection{Measurement techniques}

Conductivity of the samples was measured using a two-terminal ac technique
employing a 1211-ITHACO current preamplifier and a PAR-124A lock-in amplifier.
Most measurements were performed with the samples immersed in liquid helium at
T$\approx$4.1K held by a 100 liters storage-dewar. This allowed up to two
months measurements on a given sample while keeping it cold (and in the dark).
These conditions are essential for measurements where extended times of
relaxation processes are required at a constant temperature.

The gate-sample voltage (to be referred to as V$_{\text{g}}$ in this work) in
the field-effect measurements was controlled by the potential difference
across a 10$\mu$F capacitor charged with a constant current. The range of
V$_{\text{g}}$ used in this study reached in some cases $\pm$40V which is
equivalent to the $\pm$20V used in previous work \cite{15} where the
gate-sample separation was 0.5$\mu$m of as compared with the 1$\mu$m
SiO$_{\text{2}}$ spacer used here.

The ac voltage bias in conductivity measurements was small enough to ensure
near-ohmic conditions (except for the current-voltage plots). Optical
excitations in this work were accomplished by exposing the sample to an AlGaAs
diode operating at $\approx$0.88$\pm$0.05$\mu$m mounted on the sample-stage
$\approx$10-15mm from the sample. The diode was energized by a
computer-controlled Keithley 220 current-source.

\section{Results and discussion}

\subsection{Field-effect measurements}

Conductance versus gate-voltage G(V$_{\text{g}}$) traces typical for diffusive
GeBi$_{\text{x}}$Te$_{\text{y}}$ samples are shown in Fig.4. The sign of
$\partial$G(V$_{\text{g}}$)/$\partial$V$_{\text{g}}$ that characterize the
field-effect of these weakly-disordered samples is consistent with
band-structure calculations for GeTe and similar compounds \cite{16}. These
theoretical models account for the p-type conduction in the material, in
agreement with the sign of the Hall-Effect mentioned above. The Fermi-energy
in this scenario appears at the top of the valence-band (see inset to Fig.4).
Sweeping V$_{\text{g}}$ changes the position of the Fermi-energy and the
associated change in the conductance reflects the energy dependence of
$\partial$n$/\partial\mu$, the thermodynamic density-of-states (DOS). The
measured G(V$_{\text{g}}$) is a convoluted outcome of $\partial$n$/\partial
\mu$($\varepsilon$) and D($\varepsilon$) - the energy-dependent
diffusion-constant. For all GeBi$_{\text{x}}$Te$_{\text{y}}$ samples reported
here, $\partial$D/$\partial\varepsilon$%
$>$%
0 and is larger with larger R$_{\square}$ as evidenced by the temperature
dependence of their resistance; the resistance ratio R(4K)/R(300K) is: 1.23,
2.37, and 3.43 for the films with R$_{\square}$=6.5k$\Omega$, 32k$\Omega$, and
110k$\Omega$ in Fig.4 respectively. This explains the reason for the a general
trend in the field-effect data of disordered conductors; $\partial
$G(V$_{\text{g}}$)/$\partial$V$_{\text{g}}$ is a \textit{monotonously
increasing function of }R$_{\square}$.%
\begin{figure}[ptb]%
\centering
\includegraphics[
height=2.1421in,
width=3.0381in
]%
{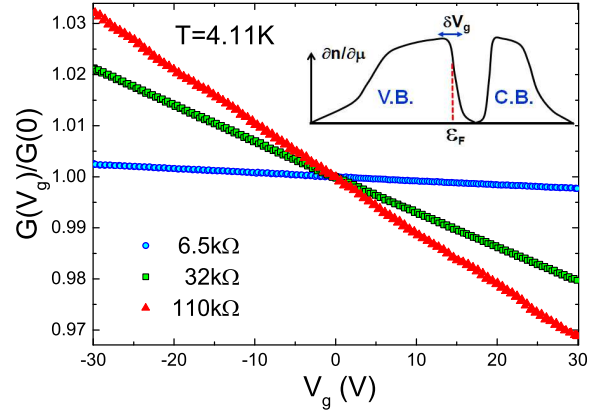}%
\caption{G(V$_{\text{g}}$) traces for three GeBi$_{\text{x}}$Te$_{\text{y}}$
films showing nearly perfect linear dependence on the gate-voltage
V$_{\text{g}}$ with a slope that increases with R$_{\square}$. The inset is a
schematic depiction of the band structure of the material and the position of
the Fermi-energy.}%
\end{figure}

Another feature appears in the G(V$_{\text{g}}$) traces of GeBi$_{\text{x}}%
$Te$_{\text{y}}$ samples that have large enough disorder. This usually
requires that the sheet-resistance of the films R$_{\square}$ be considerably
larger than h/e$^{\text{2}}\approx$25k$\Omega$. Figure 5 shows G(V$_{\text{g}%
}$) for a $\approx$50M$\Omega$ film that was taken under the same conditions
(temperature, sweep-rate) as the samples in Fig.4. The new feature is a local
depression of G(V$_{\text{g}}$) centered, in this case, at V$_{\text{g}}$=0V.
This feature is the memory-dip, a characteristic signature of the
electron-glass \cite{17,18,19,20,21}. It is believed to be a modulation of
g($\varepsilon$), the \textit{single-particle} DOS resulting from
inter-particle correlations \cite{22,23,24,25}.%

\begin{figure}[ptb]%
\centering
\includegraphics[
height=2.2909in,
width=3.039in
]%
{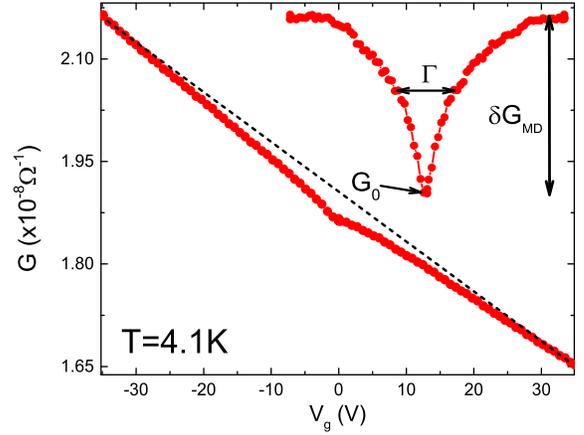}%
\caption{Conductance versus gate-voltage for a GeBi$_{\text{x}}$Te$_{\text{y}%
}$ film with R$_{\square}$ =53M$\Omega$.This G(V$_{\text{g}}$) trace was taken
24 hours after the sample was allowed to equilibrate under V$_{\text{g}}$=0V.
The dashed line stands for the thermodynamic density-of-states which,
consistent with the data in Fig.4 is assumed to be linear. The inset to the
figure shows the memory-dip of this sample obtained by subtracting the
thermodynamic DOS (the linear part of G(V$_{\text{g}}$) taken for this sample.
The memory-dip plot is marked with arrows to define its typical width $\Gamma
$, its magnitude $\delta$G$_{\text{MD}}$ and the "equilibrium" value of the
conductance, G$_{\text{0}}$. }%
\end{figure}

A modulation of g($\varepsilon$) due to correlation effects occurs in
disordered conductors even in the diffusive regime where field-effect
measurements would reveal just the thermodynamic component $\partial
$n$/\partial\mu$. The contribution of a single-particle DOS may be reflected
in tunneling or photo-emission processes but it is not expected to show up in
a thermodynamic measurement such as field-effect \cite{26}. This however is
because V$_{\text{g}}$ is usually swept sufficiently slow to allow the
electronic system to be in equilibrium. When the equilibration time of the
system becomes longer than the inverse sweep-rate (a condition that is met in
the electron-glass regime of highly disordered Anderson insulators \cite{27}),
the underlying g($\varepsilon$) would manifest itself in the G(V$_{\text{g}}$)
plot. The shape of this modulation, and its dependence on temperature, have
been discussed by Lebanon, and M\"{u}ller \cite{23}. The magnitude of the
memory-dip (defined in Fig.5), that will be referred to in this paper as
`visibility', depends therefore on how fast the gate-voltage is sweet. This
was demonstrated in field-effect experiments on crystalline indium-oxide
\cite{12}. The slope of the thermodynamic component of G(V$_{\text{g}}$), on
the other hand, is independent of $\partial$V$_{\text{g}}$/$\partial$t
\cite{12}.

The relative magnitude of the memory-dip $\delta$G$_{\text{MD}}$/G$_{\text{0}%
}$, increases with R$_{\square}$ as illustrated in Fig.6 for the
GeBi$_{\text{x}}$Te$_{\text{y}}$ films studied in this work. The average slope
of G(V$_{\text{g}}$) for this samples is plotted on the same graph for
comparison.
\begin{figure}[ptb]%
\centering
\includegraphics[
height=2.1958in,
width=3.039in
]%
{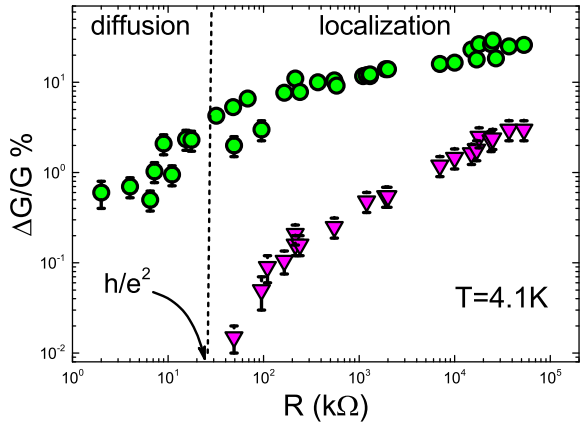}%
\caption{Comparing the magnitude of the memory-dip (triangles) and slope of
the thermodynamic density-of-states (circles) for several GeBi$_{\text{x}}%
$Te$_{\text{y}}$ films. For the latter the value for $\Delta$G/G is based on:
G(-30V)/G(+30V). }%
\end{figure}

The qualitative features of these data are remarkably similar to those
obtained on the half-a-dozen previously studied electron glasses \cite{27}. In
particular, they all show fast decline of the memory-dip magnitude as the
diffusive regime is approached from the strongly-localized side, and a similar
functional dependence of $\delta$G$_{\text{MD}}$/G$_{\text{0}}$ on
R$_{\square}$. This re-enforces the conclusion that strong-localization is a
pre-requisite for observing electron-glass effects.

Another feature that appears to be common to electron-glasses is the
dependence of their memory-dip width $\Gamma$ on the carrier-concentration
\textit{N}. A comparison of the memory-dip width and the carrier-concentration
for the three germanium-telluride compounds studied to date is given in Fig.7.
This shows a monotonous increase of $\Gamma$ with the inter-particle
separation \textit{N}$^{\text{-1/3}}$. A similar dependence was found in the
early work on amorphous indium-oxide \cite{28} which hitherto was the only
system where $\Gamma$(\textit{N}) was studied over a meaningful range of
\textit{N}. It would be interesting to be able to extend the range of the
carrier-concentration in the GeTe system towards lower \textit{N} values to
see whether the relaxation time becomes very short below a certain
concentration as it does in In$_{\text{x}}$O \cite{28}.%
\begin{figure}[ptb]%
\centering
\includegraphics[
height=2.2598in,
width=3.039in
]%
{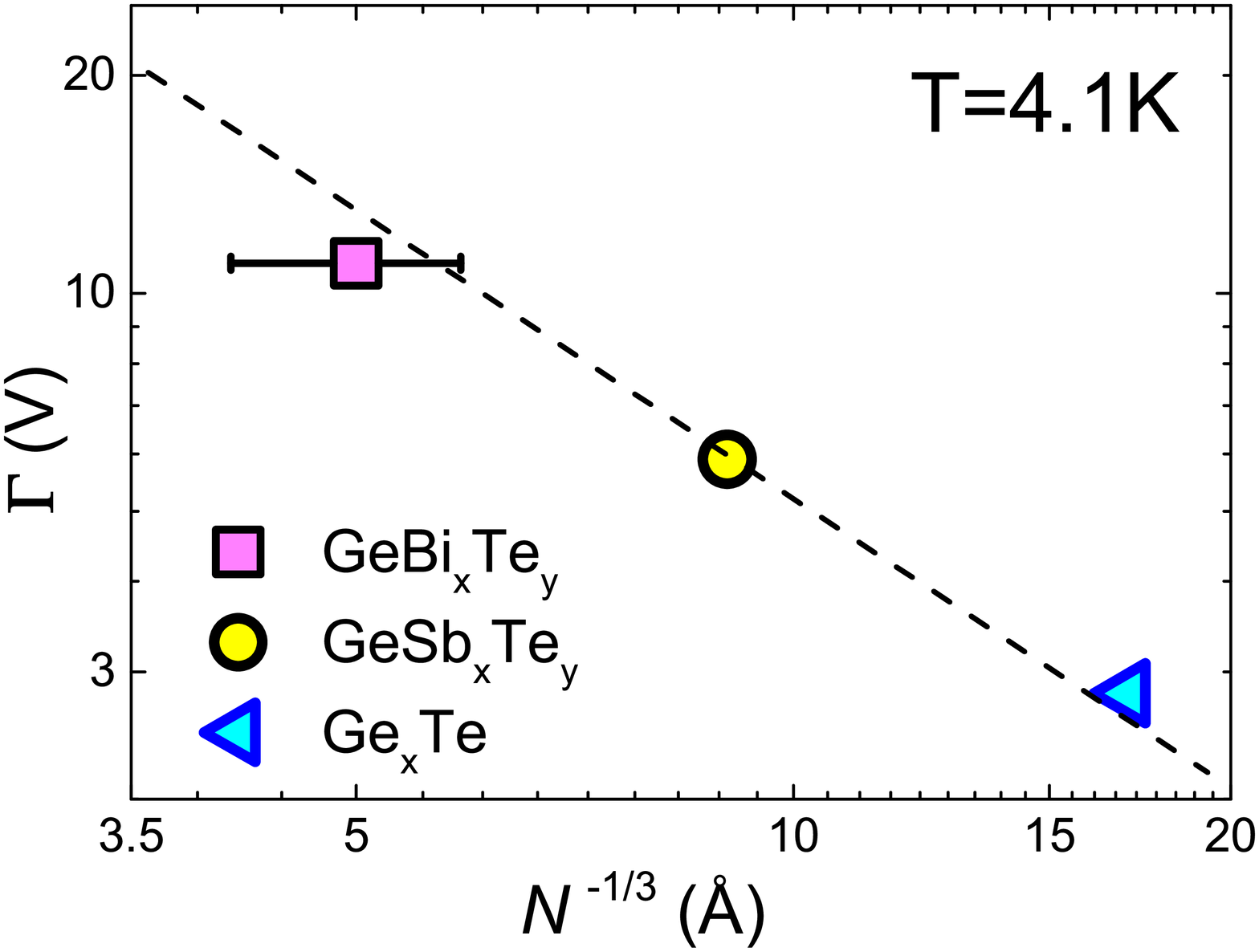}%
\caption{The width of the memory-dip $\Gamma$ (defined in Fig.5) for three
germanium-telluride alloys as function of their carrier-concentration
\textit{N}. The value for $\Gamma$(defined in the inset to Fig.5) is based on
G(V$_{\text{g}}$) measurements at T=4.1K and a SiO$_{\text{2}}$ layer of
0.5$\mu$m separating the sample for the gate.}%
\end{figure}

Quantitatively however there are some features that distinguish the
GeBi$_{\text{x}}$Te$_{\text{y}}$ samples from previously studied
electron-glasses. In particular, the magnitude of the memory-dip $\delta
$G$_{\text{MD}}$/G$_{\text{0}}$ is relatively small when compared with the
magnitude of the slope of the thermodynamic density-of-states $\partial
$n$/\partial\mu$ for a given R$_{\square}$ (Fig.6). This is mainly due to the
steep dependence of $\partial$n$/\partial\mu$ on energy of this material. This
makes it harder to observe the memory-dip in the G(V$_{\text{g}}$) traces than
in e.g., beryllium \cite{29} that has a similar magnitude of $\delta
$G$_{\text{MD}}$/G$_{\text{0}}$ for the same R$_{\square}$. The latter is
still considerably smaller than exhibited by electron-glasses with lower
carrier-concentrations \textit{N}, In$_{\text{2}}$O$_{\text{3-x}}$ and
In$_{\text{x}}$O (with N%
$<$%
10$^{\text{20}}$cm$^{\text{-3}}$), a point that will be addressed later in
this paper.

As is clear from Fig.6, a memory-dip could be observed only in strongly
insulating samples. The conductance G of these samples, like all other
electron-glasses, exhibit exponential temperature dependence of conductivity.
Temperature dependence of G is shown in Fig.8 for two of the studied
GeBi$_{\text{x}}$Te$_{\text{y}}$ samples.

It is interesting to compare these G(T) with the respective data of another
electron-glass, In$_{\text{2}}$O$_{\text{3-x}}$ which has structural features
that are similar to GeBi$_{\text{x}}$Te$_{\text{y}}$. Both show a mosaic,
polycrystalline structure. Insulating In$_{\text{2}}$O$_{\text{3-x}}$films
with comparable value of R$_{\square}$ exhibit a behavior consistent with
Mott-type hopping in two-dimensions; G(T)$\propto$exp[-(T$_{\text{0}}%
$/T)$^{\text{1/3}}$] \cite{30} rather than the simple activation G(T)$\propto
$exp[-T$_{\text{0}}$/T] shown by GeBi$_{\text{x}}$Te$_{\text{y}}$ over the
same temperature range (Fig.8). As noted above, a pre-condition for showing
electron-glass properties is strong-localization, and one of the
characteristic features of this state is an exponential G(T). However, the
specific form of this function, and the structural details of the system
appear to be unimportant. In other words, the features of electron-glasses may
not be identified by the specific form their G(T) takes. An exponential G(T)
whether simple activation or of the stretched-exponential kind is a necessary
condition but, for observing the memory-dip in field-effect measurement other
requirements have to be met, in particular strong disorder \cite{27}.

GeBi$_{\text{x}}$Te$_{\text{y}}$ and In$_{\text{2}}$O$_{\text{3-x}}$ differ
also in their sensitivity to applied electric fields; non-ohmicity in hopping
systems often sets in at rather small fields at liquid helium temperatures
\cite{31,32,33,34} and the present system is no exception; R$_{\square}$ as
function of the applied voltage V are plotted in Fig.9 for four
GeBi$_{\text{x}}$Te$_{\text{y}}$ samples with different degrees of disorder
clearly show the common trend.%
\begin{figure}[ptb]%
\centering
\includegraphics[
height=2.3817in,
width=3.039in
]%
{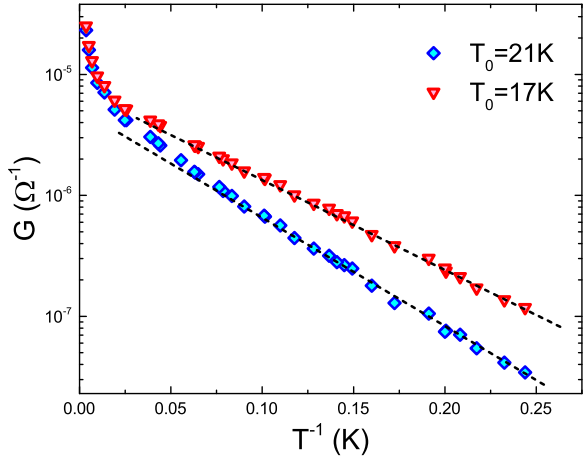}%
\caption{Conductance versus temperature for two GeBi$_{\text{x}}$%
Te$_{\text{y}}$ samples used in the study with thickness of 45\AA \ and
R$_{\square}$=8.5M$\Omega$ and R$_{\square}$=30M$\Omega$ at T=4.1K. Samples
are labeled by their activation-energies taken from the respective
R(T)$\propto\exp$[T$_{\text{0}}$/T] data.}%
\end{figure}
\begin{figure}[ptb]%
\centering
\includegraphics[
height=2.4924in,
width=3.039in
]%
{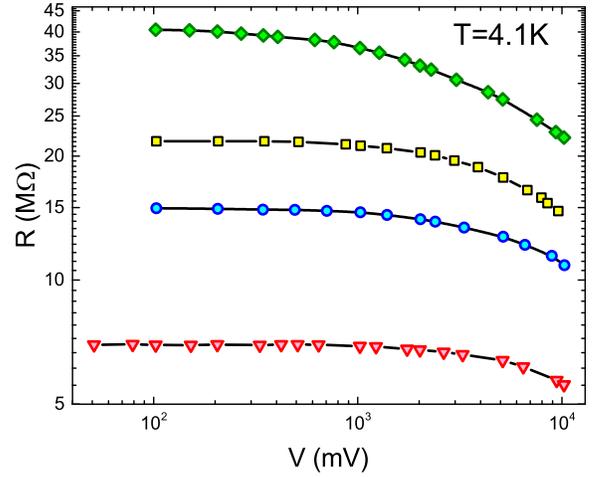}%
\caption{The dependence of sample resistance on the longitudinal voltage for
GeBi$_{\text{x}}$Te$_{\text{y}}$ films with different disorder. Samples shown
had a common geometry of 0.5mm length and 0.5mm width. }%
\end{figure}

Susceptibility to non-ohmic behavior is more conspicuous in In$_{\text{2}}%
$O$_{\text{3-x}}$ than in the GeBi$_{\text{x}}$Te$_{\text{y}}$ films: To
illustrate, In$_{\text{2}}$O$_{\text{3-x}}$ sample with R$_{\square}%
$=12.5M$\Omega$ decreased by $\approx$90\% upon applying a field of 10V/cm at
T=4.1K \cite{37} while under the same conditions the drop of R$_{\square}$ is
a mere 2.4\% for the GeBi$_{\text{x}}$Te$_{\text{y}}$ film with R$_{\square}%
$=15M$\Omega$ (Fig.9).

Non-ohmicity deep in the hopping regime is mainly caused by a field-assisted
mechanism \cite{31,32,33,34} but Joule-heating is an accompanying factor; for
a given applied field the relative importance of heating would become larger
when the resistance is smaller. The higher sensitivity of In$_{\text{2}}%
$O$_{\text{3-x}}$ to the applied field is related to the compounded effect of
both mechanisms. As a low \textit{N} system the hopping-length for a given
temperature and resistance is likely longer than in the GeBi$_{\text{x}}%
$Te$_{\text{y}}$ system which makes the field-assisted mechanism stronger for
a comparable field strength. The heating sensitivity of the conductance
involves (among other things) the electron-phonon inelastic-scattering rate.
This is likely to be more important in In$_{\text{2}}$O$_{\text{3-x}}$ due to
its unusually large Debye temperature \cite{36}. In any case, the larger range
of applied voltages over which linear-response condition may be maintained
using GeBi$_{\text{x}}$Te$_{\text{y}}$ samples should be a convenient feature
for experimental study of hopping conductivity.

There is another feature of GeBi$_{\text{x}}$Te$_{\text{y}}$ that may make it
a versatile test-bed for a variety of non-equilibrium phenomena;
GeBi$_{\text{x}}$Te$_{\text{y}}$ films exhibit pronounced response to
photo-excitation. Just like it was observed in GeSb$_{\text{x}}$Te$_{\text{y}%
}$ films \cite{9,11}, exposure to infrared illumination increased the samples
conductance by a certain $\Delta$G which persisted long after the light source
was turned off. The experimental protocol used for optical excitation is
illustrated in Fig.10. The experiment typically begins $\approx$24 hours after
the sample is cooled-down to 4.1K by recording G(t) for 1-2 minutes to
establish a baseline, near-equilibrium conductance G$_{0}$. The IR source was
then turned on for 3 seconds and then turned off while G(t) continues to be
measured.%
\begin{figure}[ptb]%
\centering
\includegraphics[
height=2.1785in,
width=3.039in
]%
{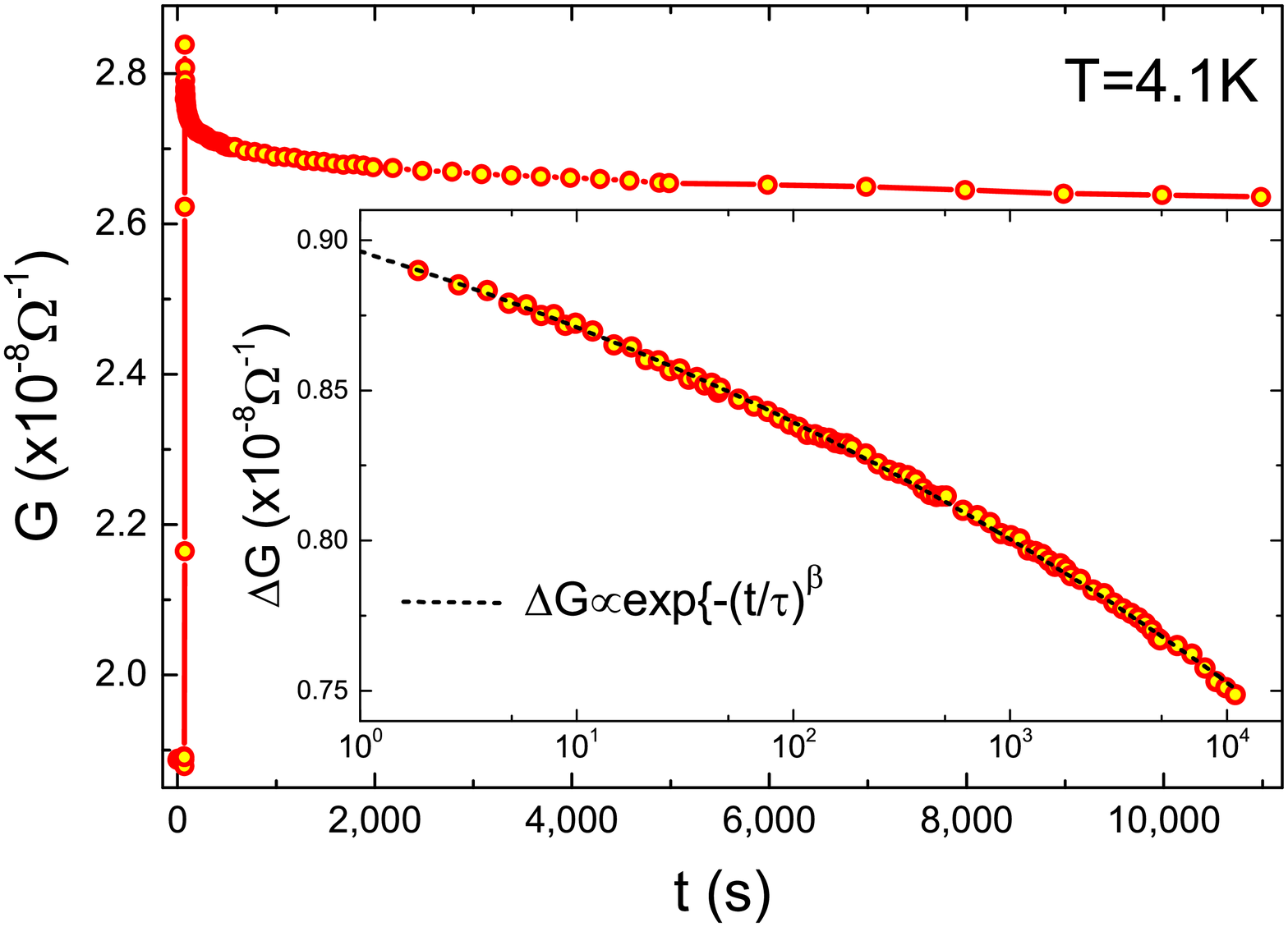}%
\caption{Conductance evolution during a typical optical excitation protocol.
The sample is a 43\AA \ thick GeBi$_{\text{x}}$Te$_{\text{y}}$ with
equilibrium R$_{\square}$=53M$\Omega$ at 4.1K. It is exposed to the infrared
LED operated with 1mA for 3s at a distance of $\approx$9mm from the sample.
The conductance is monitored for 11,000s after the LED is turned off during
which time it decays slowly with a stretched-exponential law as depicted in
the inset. A fit to $\Delta$G(t)$\propto$exp[-(t/$\tau$)$^{\beta}$] with
$\tau$=1.3x10$^{\text{9}}$s and $\beta$=0.11 is shown in the inset as a dashed
line. }%
\end{figure}

The figure depicts the conductance dependence during excitation and the
ensuing relaxation following its termination. The latter has a characteristic
time dependence that fits a stretched-exponent law:%

\begin{equation}
\text{G(t)}\propto\text{exp[-(t/}\tau\text{)}^{\beta}\text{]}%
\end{equation}
(see inset to Fig.10). This is a manifestation of persistent photoconductivity
(PPC), a phenomenon that has been frequently observed in many semiconductors
\cite{3,4,5,6,7,8} and recently found in the compound GeSb$_{\text{x}}%
$Te$_{\text{y}}$ \cite{9,11}.

In general, the relative change of the conductance due to the brief infrared
exposure $\delta$G$_{\text{IR}}$/G$_{\text{0}}$ increases with R$_{\square}$
as shown in Fig.11.%
\begin{figure}[ptb]%
\centering
\includegraphics[
height=2.1811in,
width=3.0381in
]%
{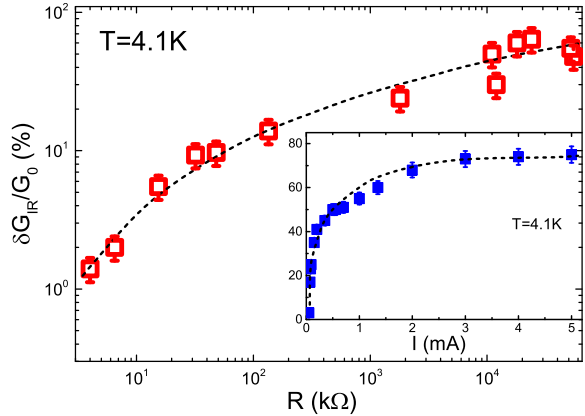}%
\caption{The relative increase of conductance due to a brief infrared exposure
$\delta$G$_{\text{IR}}$/G$_{\text{0}}$ plotted as function of the sheet
resistance of the sample. G$_{\text{0}}$ is the equilibrium value of the
conductance and $\delta$G$_{\text{IR}}$ is arbitrarily chosen as $\delta
$G$_{\text{IR}}$=G(t=3x10$^{\text{3}}$)-G$_{\text{0}}$, The inset shows the
dependence of $\delta$G$_{\text{IR}}$/G$_{\text{0}}$ on the infrared intensity
for a typical sample. Dashed lines are guides for the eye.}%
\end{figure}

The same energy-flux (1mA current through the infrared diode for 3s) was
delivered to all samples in the series with sheet-resistances of 4k$\Omega$ to
55M$\Omega$. This value was chosen after studying the dependence of the effect
on the excitation current shown in the inset to Fig.11. The excitation by 1mA
was chosen as a compromise between achieving an appreciable $\delta
$G$_{\text{IR}}$ while minimizing heating and associated detrimental effects
to the electron-glass state which may coexist with the PPC when the system is
in the strongly-localized regime.

The detailed behavior of the PPC in the GeBi$_{\text{x}}$Te$_{\text{y}}$ films
turns out to be similar to that of GeSb$_{\text{x}}$Te$_{\text{y}}$
\cite{9,11}; in both systems PPC is observable in samples that are in the
diffusive regime as well as in strongly-localized samples. The relaxation law
associated with PPC is essentially the same for a diffusive sample (Fig.12)
and for a strongly-localized sample (Fig.13). The latter is in the
electron-glass phase and exhibits a well-developed memory-dip whereas the
G(V$_{\text{g}}$) the diffusive sample reveal only the thermodynamic DOS in
the field-effect scan.%
\begin{figure}[ptb]%
\centering
\includegraphics[
height=2.0764in,
width=2.919in
]%
{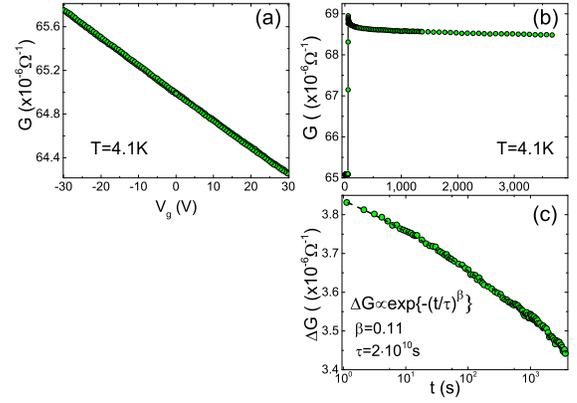}%
\caption{Data for a field-effect measurement (a) and infrared excitation
protocol (b and c) for a diffusive GeBi$_{\text{x}}$Te$_{\text{y}}$ sample
with thickness 50\AA \ and R$_{\square}$=15.4k$\Omega$ at T=4.1K. Note the
linear G(V$_{\text{g}}$) trace in (a) consistent with diffusive behavior.
Plate (c) illustrates the fit to the relaxation law where the time origin is
taken to coincide with the turning off of the infrared LED.}%
\end{figure}
\begin{figure}[ptb]%
\centering
\includegraphics[
height=2.0465in,
width=2.919in
]%
{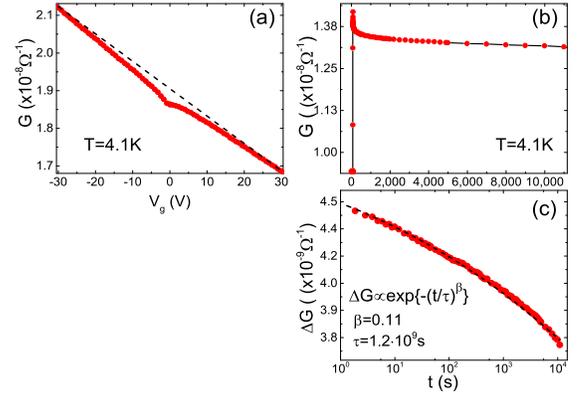}%
\caption{Data for a field-effect measurement (a) and infrared excitation
protocol (b and c) for a strongly-localized GeBi$_{\text{x}}$Te$_{\text{y}}$
sample with thickness 48\AA \ and R$_{\square}$=57M$\Omega$ at T=4.1K. Note
the appearance of a memory dip in the G(V$_{\text{g}}$) trace (a) consistent
with insulating behavior. Plate (c) illustrates the fit to the relaxation law
where the time origin is taken to coincide with the turning off of the
infrared LED.}%
\end{figure}

These two alloys are also similar in terms of magnitude and fit parameters to
the stretched exponential relaxation law. The G(t) measured following the
infrared excitation of the samples shown in Fig.11 could be fitted to Eq.1
with the same $\beta$ =0.11$\pm$0.005 as a best-fit parameter. The other fit
parameter in the stretched-exponential law is $\tau$ that is found to be of
order 10$^{\text{9}}\sec$ to 10$^{\text{10}}\sec$~in all our samples. There
was no systematic dependence of $\tau$ on the sample sheet resistance
R$_{\square}$. Similar values for $\beta$ and the relaxation-time $\tau$ were
found at these temperatures in the relaxation law of PPC in AlGa$_{\text{x}}%
$As$_{\text{1-x}}$ compounds \cite{5} and in the GeSb$_{\text{x}}%
$Te$_{\text{y}}$ films studied previously \cite{9,11}.

Applying the infrared protocol to GeMn$_{\text{x}}$Te$_{\text{y}}$ and
GeIn$_{\text{x}}$Te$_{\text{y}}$ samples resulted in a considerably lower
$\delta$G$_{\text{IR}}$/G$_{\text{0}}$ values as compared with GeSb$_{\text{x}%
}$Te$_{\text{y}}$ and the GeMn$_{\text{x}}$Te$_{\text{y}}$ samples with
similar resistances. A comparison between the response of GeMn$_{\text{x}}%
$Te$_{\text{y}}$ and GeIn$_{\text{x}}$Te$_{\text{y}}$ samples with
GeBi$_{\text{x}}$Te$_{\text{y}}$ films is shown in Fig.14.%
\begin{figure}[ptb]%
\centering
\includegraphics[
height=2.1421in,
width=3.039in
]%
{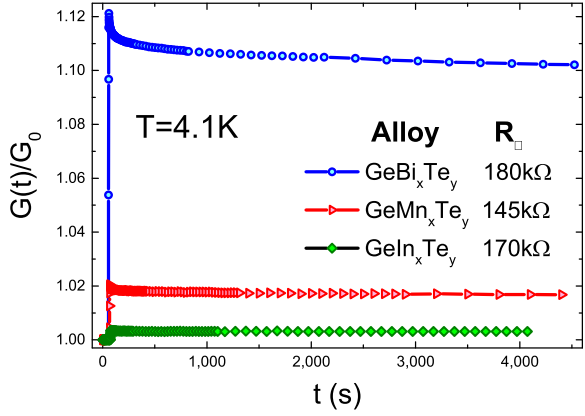}%
\caption{Comparing the result of the infrared excitation protocol for three
germanium-telluride alloys. The same protocol conditions were used in each
case as specified in Fig.10.}%
\end{figure}

The G(t) plots of the infrared protocol for these samples plotted in Fig.14
are the highest in terms of the magnitude of $\delta$G$_{\text{IR}}$ obtained
so far. The overall appearance of these relaxation curves appear to be similar
to those of the GeBi$_{\text{x}}$Te$_{\text{y}}$ and GeSb$_{\text{x}}%
$Te$_{\text{y}}$ alloys but the signal-to-noise of the G(t) data are not yet
good enough to allow for a reliable determination of $\beta$ and $\tau$ for
either GeMn$_{\text{x}}$Te$_{\text{y}}$ or GeIn$_{\text{x}}$Te$_{\text{y}}$.
In addition, there were appreciable fluctuations in the magnitude of $\delta
$G$_{\text{IR}}$/G$_{\text{0}}$ taken from different preparation batches of
these alloys. This may indicate that the PPC in GeMn$_{\text{x}}$%
Te$_{\text{y}}$ and GeIn$_{\text{x}}$Te$_{\text{y}}$ (as well as in the parent
compound GeTe) results from the spurious occurrence of a defect and that the
enhanced PPC response in the GeBi$_{\text{x}}$Te$_{\text{y}}$ and
GeSb$_{\text{x}}$Te$_{\text{y}}$ is due to a catalytic effect of incorporating
Bi or Sb in the ternary alloy.

Our persistent-photoconductivity results in these GeTe alloys resembles in
many aspects the behavior reported in compounds based on PbTe, a system that
has been extensively studied \cite{4}. In particular, the sensitivity of the
PPC magnitude to the specific chemistry of the element added to the alloy
seems to be a common feature (the dopants that yield higher PPC efficiency are
however different).

The similar dynamics of the GeBi$_{\text{x}}$Te$_{\text{y}}$ and the
GeSb$_{\text{x}}$Te$_{\text{y}}$ alloys in their PPC-state may be an important
clue for unraveling the nature of the defect responsible for the long-lived
PPC state. While the involvement of a massive defect is believed to be a key
element in most models for the PPC phenomenon \cite{4}, the basic question is
whether the relevant defect is the added element itself (Sb or Bi in our case)
or it is an indirect result of its presence in the alloy \cite{4}. Given the
substantial difference in the masses of Sb and Bi, one would argue that the
similar parameters of the relaxation dynamics in the PPC-state of
GeBi$_{\text{x}}$Te$_{\text{y}}$ and GeSb$_{\text{x}}$Te$_{\text{y}}$ as well
as the similar behavior of the undoped GeTe, favor the indirect scenario.

\subsection{Modification of the electron-glass behavior in the PPC state}

Strongly-localized samples of GeBi$_{\text{x}}$Te$_{\text{y}}$ exhibit all the
qualitative nonequilibrium features found in previously studied
electron-glasses. These are distinguishable from the PPC observed in this
system in a number of aspects. First and foremost, as mentioned above, unlike
persistent photoconductivity, the electron-glass features are not observed in
the diffusive regime of the system. Secondly, there are more ways to take the
electron-glass far from the equilibrium than just by exposure to
electromagnetic radiation; stressing the sample with longitudinal field
\cite{38}, and changing the density of carriers in the system (via a change of
the gate-voltage), are two examples for unique ways to take the system away
from equilibrium. Finally, the relaxation-law towards equilibrium of the
conductance is logarithmic with time for the electron-glass as compared with
the stretched-exponential law for PPC. Figure 15 illustrates two of the above
distinguishing features of the electron-glass.%
\begin{figure}[ptb]%
\centering
\includegraphics[
height=2.1508in,
width=3.039in
]%
{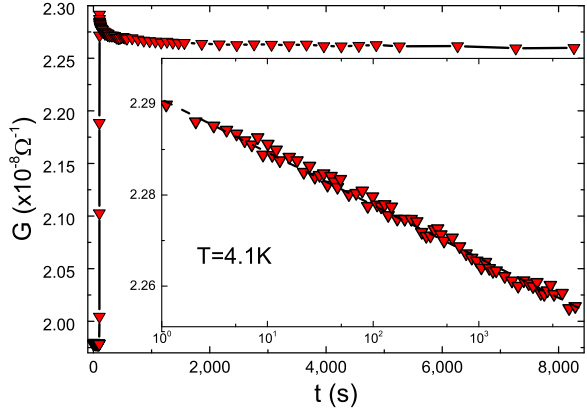}%
\caption{{}Conductance evolution during excitation protocol using a sudden
change of gate-voltage. The sample is a 43\AA \ thick GeBi$_{\text{x}}%
$Te$_{\text{y}}$ with "dark" R$_{\square}$=53M$\Omega$ at 4.1K. It was allowed
to relax under V$_{\text{g}}$=0V for 40 hours and after recording its
"equilibrium" G$_{\text{0}}$ for$\approx$10$^{\text{2}}$s, V$_{\text{g}}$ was
swept to -35V within 3.5s producing the G(t) data shown in the figure. The
inset illustrates that, after the new V$_{\text{g}}$=-35V is established, the
sample conductance decays with a logarithmic time dependence. Dashed line is
best fit for a G(t)=G(1)-a\textperiodcentered\ log(t) law. }%
\end{figure}

Some electron-glass features are significantly modified when coexisting with
persistent photoconductivity. This is not surprising; the "persistent" mode is
not a stationary-state, the concentration of charge-carriers continuously
diminishes with time, and this has consequences. It is easy to understand that
in the PPC-state the relaxation law of the electron-glass would appear
different because the background conductivity is decaying with a
stretched-exponential law. Figure 16 demonstrates such a case using the
gate-protocol to drive the electron-glass away from "equilibrium" while the
system is still relaxing from the infrared exposure applied several hours before.%

\begin{figure}[ptb]%
\centering
\includegraphics[
height=2.1871in,
width=3.039in
]%
{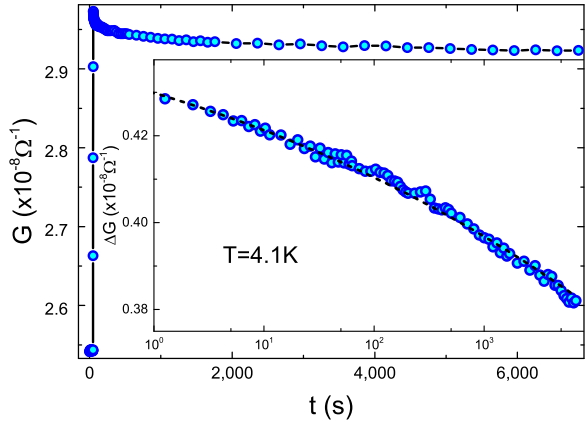}%
\caption{Conductance evolution during excitation by a gate protocol using the
same parameters as in Fig.15 except that the sample was first put in its
PPC-state as per the infrared protocol (Fig.10). The system was then allowed
to relax for three hours before changing the gate voltage. The inset
illustrates that, after the new V$_{\text{g}}$=-35V is established, the sample
conductance decays with a stretched-exponential time dependence. Dashed line
is best fit for a G(t)$\propto$exp[-(t/$\tau$)$^{\beta}$] with $\beta$=0.11
and $\tau$=3\textperiodcentered10$^{\text{10}}$s.}%
\end{figure}

The main effect of the combined relaxation in this case is a modified
relaxation-law that deviates from the log(t) dependence that characterizes the
electron-glass. The total change of $\Delta$G observed during the $\approx
$7,000 seconds, over which data are plotted in Fig.16, is comparable in
magnitude to the sum of the respective $\Delta$G's estimated from the
relaxations due to the PPC and the electron-glass (assuming that each is
present without the other). This assumption however cannot be accurate;
coexistence of the two nonequilibrium phenomena manifestly produces effects
that are not consistent with simple superposition. This is demonstrated in the
experiment described in Fig.17. The figure compares G(V$_{\text{g}}$) for a
GeBi$_{\text{x}}$Te$_{\text{y}}$ film before and after being exposed to the
infrared source yielding. a counter-intuitive result: The sample under the
PPC-state exhibits a \textit{larger} amplitude of the memory-dip than the same
film in its "dark" state.
\begin{figure}[ptb]%
\centering
\includegraphics[
height=2.444in,
width=3.039in
]%
{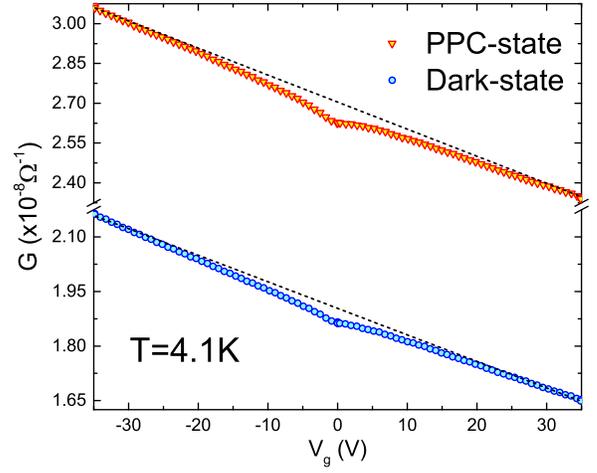}%
\caption{Field-effect G(V$_{\text{g}}$) traces for a GeBi$_{\text{x}}%
$Te$_{\text{y}}$ sample with "dark" R$_{\square}$=54M$\Omega$ taken 24 hours
after cooldown to T=4.1K (circles) and 3 hours after exposing the film to
infrared radiation at 1mA for 3s (squares). Both traces were taken with the
same sweep rates of 0.023V/s. The relative slopes $\Delta$G/G of the
thermodynamic component of G(V$_{\text{g}}$) (see Fig.6 for definition) were
$\approx$30\%$\pm$0.1\% for both curves while the magnitude of the memory-dip
$\delta$G$_{\text{MD}}$/G$_{\text{0}}$ is 2\%$\pm$0.1\% in the dark-state and
2.9\%$\pm$0.1\% in the PPC-state.}%
\end{figure}

This is not a trivial result; it is natural to expect that exposing the
electron-glass to infrared radiation would result in a \textit{diminishment}
of the memory-dip magnitude. In fact, it has been shown that above a certain
threshold of energy-flux, infrared illumination may completely erase the
memory-dip \cite{39}. The energy flux associated with initiating the PPC-state
is much smaller than used in \cite{39} but the enhancement of the memory-dip
is still an unexpected result.

A similar effect in GeSb$_{\text{x}}$Te$_{\text{y}}$ was ascribed to the
presence of a series-resistor that is reduced in the PPC-state \cite{9,11}.
This was motivated by the observation of the concomitant increase of the slope
of $\partial$n$/\partial\mu$($\varepsilon$). However, this approach was not
able to account for the magnitude of the enhancement. Besides, the lack of
change in the slope of the thermodynamic DOS in the present case [see Fig.17],
makes this line of explanation questionable anyhow (although, given the
inhomogeneous structure, part of the effect might be related to reduction of a
series resistor).

An alternative scenario to account for the enhanced memory-dip is based on the
observation that the interaction-to-disorder ratio plays a role in the
magnitude of the memory-dip and on the realization that this ratio increases
in the PPC-state relative to the dark-state.

{\small T}hat the visibility of the memory-dip is a function of the ratio of
Coulomb-interaction I$_{\text{C}}$ to disorder W should be clear; recall that
this modulation in G(V$_{\text{g}}$) has it roots in the competition between
interaction and disorder: The memory-dip vanishes when disorder is small, and
it is obviously absent altogether when interaction is turned-off. It is then
intuitively expected that the memory-dip is most conspicuous when the disorder
and interaction are comparable in magnitude.

The first point we wish to make here is that, over the range of parameters
relevant for \textit{all} the electron-glasses studied to date, I$_{\text{C}%
}\gg$W.

The magnitude of the interparticle Coulomb interaction I$_{\text{C}}$ for a
system with a given \textit{N} can be expressed by:%

\begin{equation}
\text{I}_{\text{C}}=\frac{e^{\text{2}}}{\varepsilon r}=\frac{e^{\text{2}%
}\mathit{N}^{\text{1/3}}}{\varepsilon}%
\end{equation}

where $\varepsilon$ is the dielectric constant of the medium and
$r=$\textit{N}$^{\text{-1/3}}$ is the interparticle average spacing. The
disorder required to make the system strongly-localized (the pre-condition for
being in the electron-glass phase \cite{27}) :%

\begin{equation}
\text{W}\approx\text{6.5\textperiodcentered E}_{_{\text{F}}}%
\end{equation}

and using free-electron formula this may be cast in terms of carrier-concentration:%

\begin{equation}
\text{W}\approx\text{6.5\textperiodcentered}\frac{\hslash^{\text{2}}%
\text{k}_{\text{F}}^{\text{2}}}{\text{2m*}}\text{;}~\text{and k}_{\text{F}%
}=(\text{3}\pi^{\text{2}}\mathit{N})^{\text{1/3}}%
\end{equation}

Note that all currently known electron-glasses have carrier-concentration in
the range 2x10$^{\text{19}}$cm$^{\text{-3}}\leq$\textit{N}$\leq$%
5x10$^{\text{21}}$cm$^{\text{-3}}$. As explained in \cite{27}, systems with
higher carrier-concentration than \textit{N}$\approx$10$^{\text{22}}%
$cm$^{\text{-3}}$ are difficult to strongly-localize unless by making
granular, and the disorder in systems with \textit{N}$\eqslantless
$10$^{\text{19}}$cm$^{\text{-3}}$ is too weak to reduce transition rates
sufficiently as to afford observation of the memory-dip (this is the reason
for the failure of lightly-doped semiconductors to exhibit a memory-dip
\cite{27}).

The values for W and I$_{\text{C}}$ may now be estimated using Eq.2 and Eq.4
with the parameters for indium-oxide which has been studied over the widest
range of carrier-concentration among all electron-glasses. Taking the
dielectric constant as $\varepsilon$=10, and effective mass m*=0.3m$_{\text{0}%
}$ one gets: I$_{\text{C}}$/W$\approx$0.14 and I$_{\text{C}}$/W$\approx$0.025
for In$_{\text{x}}$O samples with \textit{N}$\approx$4x10$^{\text{19}}%
$cm$^{\text{-3}}$ and $\hspace{0.01in}$\textit{N}$\approx$5x10$^{\text{21}}%
$cm$^{\text{-3}}$ respectively. These represent the two extreme limits in
\textit{N} over which electron-glasses were studied. The memory-dips
associated with these limit-\textit{N} values are plotted in Fig.18
illustrating the much larger $\delta$G$_{\text{MD}}$/G$_{\text{0}}$ exhibited
by the sample with the larger I$_{\text{C}}$/W ratio.%
\begin{figure}[ptb]%
\centering
\includegraphics[
height=2.1551in,
width=3.039in
]%
{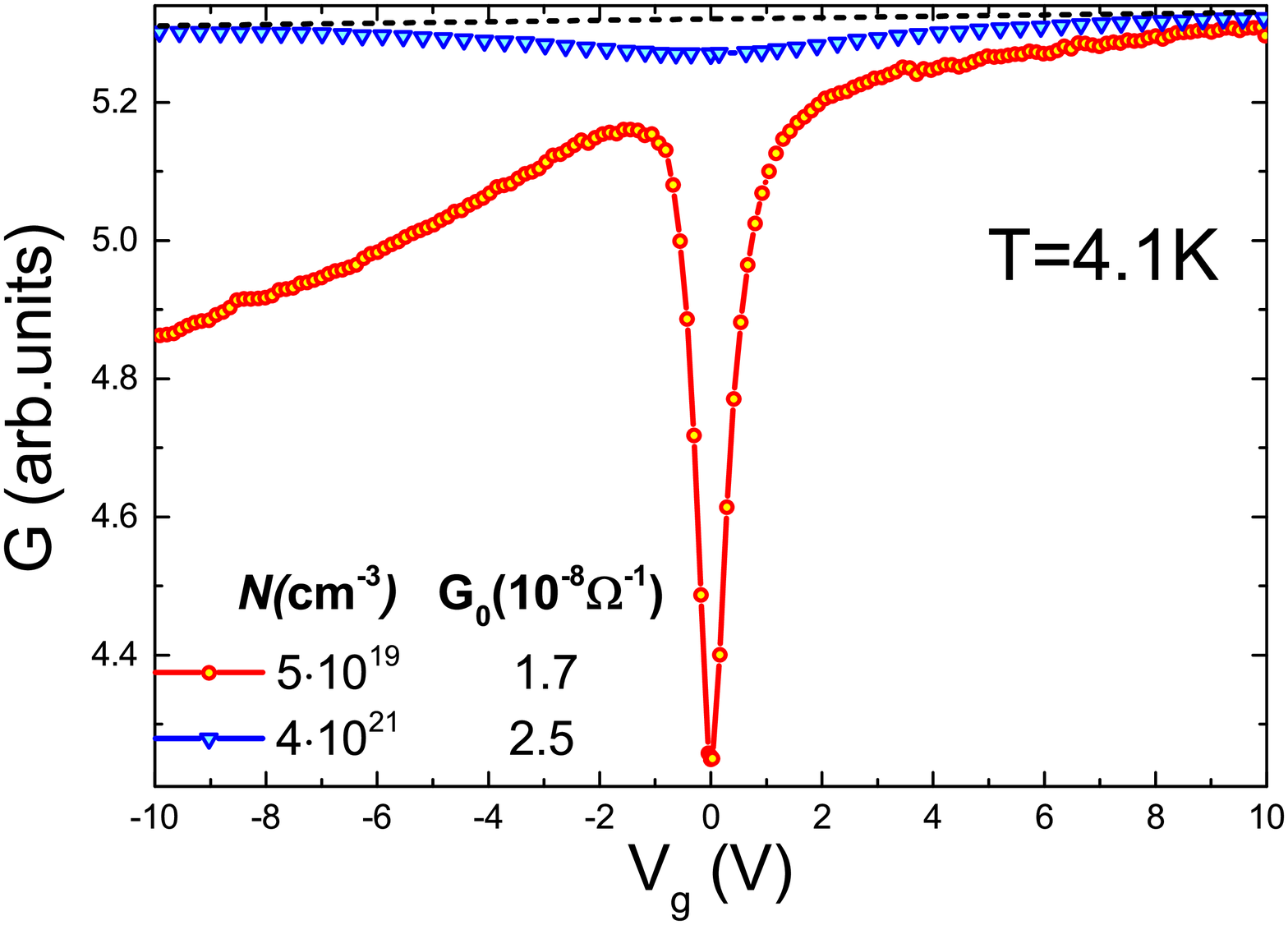}%
\caption{Field-effect sweeps comparing the memory-dip of two In$_{\text{x}}$O
samples with different values of \textit{N} (obtained by different In/O
composition) but similar R$_{\square}$ and measured under identical conditions
(sweep rate, relaxation history). These G(V$_{\text{g}}$) plots are based on
the same field-effect structures described in section II but with 0.5$\mu$m
SiO$_{\text{2}}$ spacer. Dashed line is the assumed thermal-DOS plotted to
expose the wide and shallow memory-dip of the sample with the larger
\textit{N}. The latter has $\delta$G$_{\text{MD}}$/G$_{\text{0}}\approx$1.1\%
while the low \textit{N} sample exhibits $\delta$G$_{\text{MD}}$/G$_{\text{0}%
}\approx$21\%$.$}%
\end{figure}

With these parameters for In$_{\text{x}}$O, the interaction-to-disorder ratio
I$_{\text{C}}$/W$\approx$1 would be obtained for an Anderson insulator with
carrier-concentration \textit{N}$\approx$3x10$^{\text{16}}$cm$^{\text{-3}}$.
This \textit{N} is well below the concentration where the relaxation time
becomes too short for observation of electron-glass features by standard
field-effect techniques. Note that it is the system with the lower \textit{N}
that has the larger I$_{\text{C}}$/W (and consequently larger $\delta
$G$_{\text{MD}}$/G$_{\text{0}}$). This is so because I$_{\text{C}}\propto
$\textit{N}$^{\text{1/3}}$ while for electron-glasses W$\propto$%
\textit{N}$^{\text{2/3}}$, thus I$_{\text{C}}$/W actually decreases with
\textit{N}. In other words, all other things being equal, higher
carrier-density means stronger Coulomb interaction but even stronger disorder.

Large values for $\delta$G$_{\text{MD}}$/G$_{\text{0}}$, of the order of
15-30\% (for R$_{\square}$ $\approx$10M$\Omega$ at 4K), are routinely obtained
using the crystalline version of indium-oxide (In$_{\text{2}}$O$_{\text{3-x}}%
$) that also has low carrier-concentration \textit{N}$\approx$4x10$^{\text{19}%
}$cm$^{\text{-3}}$ \cite{12}. On the other hand, electron-glasses with
\textit{N} $\geq$10$^{\text{21}}$cm$^{\text{-3}}$ typically show $\delta
$G$_{\text{MD}}$/G$_{\text{0}}$ of the order of $\approx$1\% in samples with
comparable R$_{\square}$. This is manifestly the case for the GeBi$_{\text{x}%
}$Te$_{\text{y}}$ films studied here (see Fig.6 above).

The larger $\delta$G$_{\text{MD}}$ observed in the PPC-state in both
GeBi$_{\text{x}}$Te$_{\text{y}}$ and the GeSb$_{\text{x}}$Te$_{\text{y}}$
alloys may just be another example of the dependence of the memory-dip
magnitude on I$_{\text{C}}$/W. Note that the enhanced conductivity in the
PPC-state is mainly due to higher concentration of carriers \textit{N} created
by the optical excitation. Being in the strongly-localized regime, the system
lacks metallic screening and therefore higher \textit{N} means an increase in
the interparticle-interaction I$_{\text{C}}$. It is less clear what, if any,
is the accompanying change of disorder; The defect that hinders recombination
in the PPC-state is presumed to be stabilized by a spatial-shift $\Delta$x of
an atom, and possibly a change in the local charge \cite{8}. The atomic
displacement $\Delta$x, is sub-atomic \cite{8} and thus much smaller than the
localization-length which can hardly affect the perceived ionic-disorder. The
scattering cross-section may however be different due to a change in the local
charge in the PPC-state, but this may go either way. It is then plausible to
assume that, in the PPC-state, the ratio of Coulomb-interaction to disorder is
larger than in the dark-state. In that case the enhanced $\delta$%
G$_{\text{MD}}$/G$_{\text{0}}$ in the PPC-state just follows a trend generally
obeyed by all electron-glasses with \textit{N}$\geq$3x10$^{\text{19}}%
$cm$^{\text{-3}}$.

To summarize, we presented in this paper experimental results that demonstrate
coexistence of persistent-photoconductivity and electron-glass features in the
degenerate semiconductor GeBi$_{\text{x}}$Te$_{\text{y}}$. Both phenomena
exhibit sluggish conductance relaxation albeit due to different mechanisms.
The conductance in the persistent-state is associated with the recombination
of optically-generated excess charge. The process is slow presumably due to an
energy-gap induced by a local structure re-arrangement. The different PPC
susceptibility of the system to the nature of the element added to the alloy
may help in identifying the microscopic mechanism involved.

The electron-glass dynamics, on the other hand, is associated with a change in
the carriers \textit{mobility }rather than carrier-concentration, and it is
controlled by the combined effects of quench-disorder and variety of
interaction-related mechanisms: Many-body effects, and several variations of
the orthogonality-catastrophe \cite{40,41,42} may further extend relaxation
times. These phenomena demonstrate the richness and complexity of electronic
transport in disordered and interacting quantum systems. The current study
re-affirms the notion that electron-glasses with long relaxation times are
inherent property of Anderson insulators where disorder is much larger than
the Coulomb interaction.

The enhancement of the memory-dip magnitude induced by infrared radiation lead
us to the conjecture that this feature of the electron-glass is controlled by
the ratio of interaction and disorder. This was shown to be consistent with a
number of experiments. Further experimental work needs to be done to test this
trend in different materials. The magnitude $\delta$G$_{\text{MD}}%
$/G$_{\text{0}}$ of the memory-dip is a more complicated issue to characterize
than the width $\Gamma$. At a given temperature, $\Gamma$ depends only on
carrier-concentration of the system \cite{12}. The magnitude of the
memory-dip, on the other hand, depends on many other factors; $\delta
$G$_{\text{MD}}$/G$_{\text{0}}$ depends on the rate of sweeping the
gate-voltage V$_{\text{g}}$, it depends on the time the sample equilibrated
under V$_{\text{g}}$ (history), and it depends on the R$_{\square}$ and on the
sample thickness. In addition it may vary between different preparation
batches (which suggest influence of structural details). The functional
dependence of $\delta$G$_{\text{MD}}$/G$_{\text{0}}$ on carrier-concentration
is therefore not yet established. Nevertheless, the overall trend that
$\delta$G$_{\text{MD}}$/G$_{\text{0}}$ increases when the
carrier-concentration decreases (in the regime I$_{\text{C}}\ll$W) is clear
enough. Some insight on this issue may come from numerical simulations looking
at how the I$_{\text{C}}$/W ratio affects the memory-dip magnitude using the
method of Meroz et al \cite{25}; in principle using this technique should
allow probing the regime where the interaction is larger than the disorder,
which is hard to implement experimentally.

\begin{acknowledgments}
Illuminating discussions with Ady Vaknin on the electron-glass dynamics and
with Dmitry Khokhlov on persistent-photoconductivity are gratefully
acknowledged. This research has been supported by a grant No 1030/16
administered by the Israel Academy for Sciences and Humanities.
\end{acknowledgments}


\begin{thebibliography}{99}                                                                                               %


\bibitem {1}Marcos Rigol1, Vanja Dunjko, and Maxim Olshanii, Nature,
\textbf{452}, 854 (2008).

\bibitem {2}Anatoli Polkovnikov, Krishnendu Sengupta, Alessandro Silva, and
Mukund Vengalattore, Rev. Mod. Phys. \textbf{83}, 863 (2011).

\bibitem {3}N. G. Ptitsina, G. M. Chulkova, K. S. Il'in, A. V. Sergeev, F. S.
Pochinkov, E. M. Gershenzon, and M. E. Gershenson, Phys. Rev. B \textbf{56},
10089 (1997).

\bibitem {4}B. A. Akimova,. V. Dmitriev., R. Khokhlov, L. I. Ryabova, phys.
stat. sol. (a) \textbf{137}, 9 (1993).

\bibitem {5}T. Y. Lin, H. M. Chen, M. S. Tsai, and Y. F. Chen, F. F. Fang, C.
F. Lin and G. C. Chi, Phys. Rev. B \textbf{58}, 13793 (1998).

\bibitem {6}H. J. Queisser and D. E. Theodorou, Phys. Rev. Lett., \textbf{43},
401 (1979); Jennifer Misuraca, Jelena Trbovic, Jun Lu, Jianhua Zhao, Yuzo
Ohno, Hideo Ohno, Peng Xiong, and Stephan von Moln\'{a}r, Phys. Rev. B
\textbf{82}, 125202, (2010).

\bibitem {7}P. M. Mooney, Journal of Applied Physics \textbf{67}, R1 (1990);
Z. Su and J. W. Farmer, Appl. Phys. Lett. \textbf{59} , 30 (1991); D. Jia, J.
Zhu, and B. Wu, Journal of The Electrochemical Society, \textbf{147} (1) 386 (2000).

\bibitem {8}D. V. Lang and R. A. Logan, Phys. Rev. Lett., \textbf{39,} 635
(1977); D. V. Lang, R. A. Logan, and M. Jaros, Phys. Rev. B\textbf{\ 19}, 1015
(1979). The notion of a gap created by an atom-shift, which is central to the
model, was first used by R. W. Gurney and N. F. Mott, in Trans. Faraday SOC.
\textbf{35}, 69 (1939).

\bibitem {9}Z. Ovadyahu , Phys. Rev. B. \textbf{91}, 094204 (2015).

\bibitem {10}By "intrinsic" we mean that the glassy effects appear in a given
substance \textit{independently} of the way the sample was prepared to achieve
the required parameters (resistivity at the measuring temperature,
carrier-concentration, and dimensionality). Most importantly, the system has
to exhibit a memory-dip with a width that is commensurate with the
carrier-concentration of the material.

\bibitem {11}Z. Ovadyahu, Phys. Rev. Lett., \textbf{115}, 046601 (2015).

\bibitem {12}Z. Ovadyahu, Phys. Rev. B \textbf{78}, 195120 (2008).

\bibitem {13}U. Givan and Z. Ovadyahu, Phys. Rev. B \textbf{86}, 165101 (2012).

\bibitem {14}Z. Ovadyahu, Phys. Rev. B. 94, 155151 (2016).

\bibitem {15}A. Vaknin, Z. Ovadyahu, and M. Pollak, Phys. Rev. B\textbf{\ 65},
134208 (2002).

\bibitem {16}A. H. Edwards A. C. Pineda, P. A. Schultz, M.s G. Martin, A. P.
Thompson, and H.P. Hjalmarson, C. J. Umrigar, J. Phys.: C Condens. Matter
\textbf{17,} \ L329\ \ (2005); \textit{ibid} Phys. Rev. B \textbf{73}, 045210 (2006).

\bibitem {17}J. H. Davies, P. A. Lee, and T. M. Rice, Phys. Rev. Letters,
\textbf{49}, 758 (1982).

\bibitem {18}M. Gr\"{u}newald, B. Pohlman, L. Schweitzer, and D. W\"{u}rtz, J.
Phys. C, \textbf{15}, L1153 (1982).

\bibitem {19}M. Pollak and M. Ortu\~{n}o, Sol. Energy Mater., \textbf{8}, 81
(1982); M. Pollak, Phil. Mag. B\textbf{\ 50}, 265 (1984).

\bibitem {20}G. Vignale, Phys. Rev. B\textbf{\ 36}, 8192 (1987).

\bibitem {21}Ariel Amir, Yuval Oreg, and Yoseph Imry, Annu. Rev. Condens.
Matter Phys. \textbf{2,} 235 (2011); M. Pollak, M. Ortu\~{n}o and A. Frydman,
"\textit{The Electron Glass}\textbf{",} Cambridge University Press, England (2013).

\bibitem {22}C. C. Yu, Phys. Rev. Lett., \textbf{82}, 4074 (1999).

\bibitem {23}Eran Lebanon, and Markus M\"{u}ller, Phys. Rev. B\textbf{\ 72},
174202 (2005).

\bibitem {24}R. Grempel, Europhys. Lett., \textbf{66,} 854 (2004).

\bibitem {25}Y. Meroz, Y. Oreg and Y. Imry, EPL, \textbf{105}, 37010 (2014).

\bibitem {26}P. A. Lee, Phys. Rev. B\ \textbf{26,} 5882 (1982).

\bibitem {27}Z. Ovadyahu, Phys. Rev. B. \textbf{95}, 134203 (2017).

\bibitem {28}A. Vaknin, Z. Ovadyhau, and M. Pollak, Phys. Rev. Lett.,
\textbf{81}, 669 (1998).

\bibitem {29}Z. Ovadyahu, X. M. Xiong, and P. W. Adams, Phys. Rev. B
\textbf{82}, 195404 (2010).

\bibitem {30}A. Vaknin, A. Frydman, Z. Ovadyahu, and M. Pollak, Phys. Rev. B
\textbf{54}, 13604 (1996).

\bibitem {31}N. Apsley and H.P. Hughes, Philos. Mag. \textbf{31}, 1327 (1975).

\bibitem {32}M. Pollak and I. Riess, J. Phys. \textbf{C9}, 2339 (1976).

\bibitem {33}R. M. Hill, Philos. Mag. \textbf{24}, 1307 (1971).

\bibitem {34}B. I. Shklovskii, Fiz. Tekh. Poluprovodn. \textbf{6}, 2335 (1972)
[Sov.Phys. Semicond. \textbf{6}, 1964 (1973)].

\bibitem {35}Z. Ovadyahu, Phys. Rev. Lett., \textbf{108}, 156602 (2012).

\bibitem {36}I. Schwartz, S. Shaft, A. Moalem and Z. Ovadyahu, Phil. Mag. B
\textbf{50}, 221 (1984).

\bibitem {37}Z. Ovadyahu, Phys. Rev. Lett., \textbf{108}, 156602 (2012).

\bibitem {38}V. Orlyanchik, and Z. Ovadyahu, Phys. Rev. Lett., \textbf{92},
066801 (2004).

\bibitem {39}V. Orlyanchik, A.Vaknin, Z. Ovadyahu, and M. Pollak, Phys. Stat.
Sol., B\textbf{ 230}, 61 (2002).

\bibitem {40}Z. Ovadyahu, Phys. Rev. Lett., \textbf{99}, 226603 (2007).

\bibitem {41}Vedika Khemani, Rahul Nandkishore, and S. L. Sondhi, Nature
Physics, \textbf{11}, 560 (2015).

\bibitem {42}D. L. Deng, J. H. Pixley, X Li, S. D. Sarma, Phys. Rev. B.
\textbf{92}, 220201(R) (2015).
\end{thebibliography}
\end{document}